# Design of Quantum Annealing Machine for Prime Factoring

M. Maezawa, K. Imafuku, M. Hidaka, H. Koike, S. Kawabata
National Institute of Advanced Industrial Science and Technology
Tsukuba, Ibaraki 305-8565, Japan
masaaki.maezawa@aist.go.jp

*Abstract*—We propose a prime factoring machine operated in a frame work of quantum annealing (QA). The idea is inverse operation of a quantum-mechanically reversible multiplier implemented with QA-based Boolean logic circuits. We designed the QA machine on an application-specific-annealing-computing architecture which efficiently increases available hardware budgets at the cost of restricted functionality. The circuits are to be implemented and fabricated by using superconducting integrated circuit technology. We propose a three-dimensional packaging scheme of a qubit-chip / interposer / package-substrate structure for realizing practically-large scale QA systems.

## I. Introduction

Since Shor's discovery of quantum algorithms [1], prime factoring of large numbers has been the biggest game in the field of quantum computing [2]-[4] because its difficulty guarantees the security of RSA cryptosystem [5]. On the other hand, a new type of quantum computing, *i.e.*, quantum annealing (QA) was proposed by Kadowaki and Nishimori [6], and has been pursued and developed mainly by D-Wave Systems in the past decade [7]-[11]. While D-Wave-type QA machines are considered promising for solving optimization and sampling problems, they can also solve various problems expressed in forms of combinational logic on the basis of Ising-model implementation of Boolean logic [12]. In this paper we present design of a superconducting QA-based factoring machine which is basically the same as a Boolean multiplier operated in the inverse direction. Our plan toward a practical-scale factoring machine is discussed from concept to technology.

## II. Quantum Annealing for Factoring

The QA process is reversible because its time evolution is unitary. Fig. 1 shows an operation principle of a factoring machine which is a QA multiplier operated inversely [12]. The QA machine, multiplier and/or factorizer, is designed to have the minimum energy when the boundary states satisfy $P = M \times N$. By setting $M$ and $N$, the product $P = M \times N$ is obtained at the energy minimum after the annealing process. In a similar way, by setting $P$, one of the possible combinations of the factors $M$ and $N$ is probabilistically obtained.

The QA factoring machine is simply designed as a multiplier composed of Boolean logic gates in the same way as classical digital circuits. But differently from the classical circuits, the logic gates consist of qubits and inter-qubit couplers to be operated quantum mechanically. The QA implementation of the Boolean logic gates is expressed by Hamiltonian of Ising model,

$$H = \sum_i h_i \sigma_i + \sum_{i,j} J_{ij} \sigma_i \sigma_j \quad , \qquad (1)$$

where $\sigma_i$ is an $i$-th qubit spin, $h_i$ is a qubit bias energy, and $J_{ij}$ is a coupling energy between qubits [12]. For instance, a QA NOR gate has $h_1 = h_2 = 0.5$, $h_3 = 1$, $J_{12} = 0.5$ and $J_{13} = J_{23} = 1$ (Fig. 2(a)). An arbitrary Boolean logic circuit is constructed by accordingly connecting the QA logic gates with $J_{ij} = -1$. As shown in Fig. 2(b), a QA half adder is composed of three QA NOR gates, $Q_1$-$Q_2$-$Q_3$, $Q_4$-$Q_5$-$Q_6$ and $Q_7$-$Q_8$-$Q_9$. Here, the $Q_3$-$Q_8$ and $Q_6$-$Q_7$ couplings with $J_{38} = J_{67} = -1$ plainly interconnect the qubits whereas the $Q_1$-$Q_4$ and $Q_2$-$Q_5$ couplings with $J_{14} = J_{25} = 1$ provide not only interconnection but also NOT operation. Assuming the straightforward implementation like a half adder in Fig. 2(b), we roughly estimate that $25p^2$ qubits carry out factoring of a $p$-bit number.

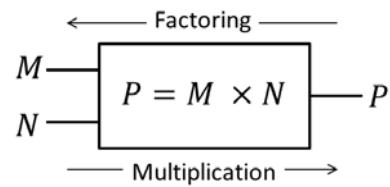

Fig. 1. Concept of factoring based on the reversivility of QA process.

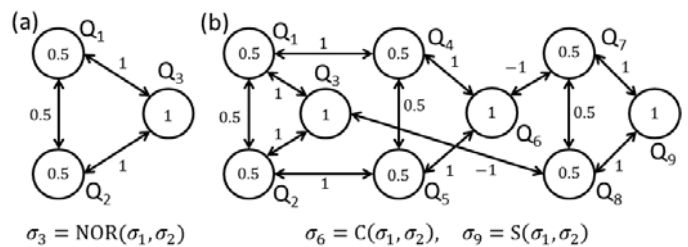

Fig. 2. Examples of QA implementations of Boolean logic circuits. (a) NOR and (b) half adder. Circles denote qubits $Q_i$ with spin $\sigma_i$ and numbers inside are the coefficients $h_i$. Double sided arrows denote inter-qubit couplings between $Q_i$ and $Q_j$, and numbers aside are the coefficients $J_{ij}$.

## III. ARCHITECTURE AND IMPLEMENTATION

The QA machine for factoring is to be realized using superconducting integrated circuit technology that is the most advanced one among solid-state quantum circuit technologies [13][14]. However, a strong constraint in the architecture layer is the limited hardware budget associated with the immaturity of the superconducting circuit manufacturing. Even state-of-the-art Nb-based technologies allow the integration of $10^4$ to $10^5$ Josephson junctions per chip [15][16], which is four to five orders of magnitude lower than those in today's semiconductor circuits. Our strategy is restriction of the functionality: we intend to develop a QA machine only for factoring applications. The target QA machine has a fixed structure of the qubit network with built-in parameters, $h_i$ and $J_{ij}$, specialized for the factoring operation. The special-purpose approach, which we name an Application Specific Annealing Computing (ASAC) architecture, efficiently reduces the hardware overhead as well as the cost and time for development. Although the limited faculty limits the applications, we expect a complementary role with general purpose QA machines like D-Wave machine.

The circuits are implemented with Josephson junctions and superconducting inductors. Following D-Wave [8]-[10], we employ a superconducting flux qubit whose potential is tunable with transvers fields. In contrast with D-Wave machine, the inter-qubit coupler is a passive transformer with fixed mutual inductance. Fig. 3 shows a simplified schematic of the qubit which is equipped with eight connection ports, C1 to C8, for connecting other qubits, control circuits and a readout circuit. The qubit bias $h_i$ is set with a current $I_x$. The coupling strength $J_{ij}$ is fixed but discretely changed by selecting the built-in number of the connection ports for coupling. The polarity of $J_{ij}$, positive or negative, is also selectable by changing the connection polarity of the ports. The transvers field is induced by an external input current $I_t$ for annealing. The circuit parameters such as the junction critical currents and inductances are under optimization.

Fig. 4 shows a block diagram of a QA NOR gate consisting of three superconducting qubits, $Q_1$, $Q_2$ and $Q_3$. Blank ports are used for the inter-gate connections, fine bias control and readout.

## IV. TECHNOLOGY INTEGRATION

As discussed at the beginning of Section III, the number of available Josephson junctions or superconducting qubits per chip is strictly limited. At present, for instance, the existing largest-scale QA machine, D-Wave 2000Q, has at most 2000 qubits, which consists of a single QA processor chip [7]. Multi-chip module (MCM) packaging is therefore an essential technology for realizing a practically-large-scale QA machine for solving practical problems. Another big challenge in the technology layer is how to meet the different technology requirements for different types, quantum and classical, of circuits. On one hand, low-$J_c$ Josephson junctions in submicron sizes are necessary for the quantum circuits, qubits and couplers, to reduce both subgap leakage currents and parasitic capacitance for achieving long decoherence times. On the other hand, higher-$J_c$ junctions in moderate sizes are suitable for fabricating the classical circuits for control and readout of the qubits. Our decision is separation of the fabrication processes: the qubits and couplers are integrated on reasonably small-size chips for maintaining sufficient yields of working chips; the auxiliary control and readout circuits are built in large-size interposers for facilitating the MCM packaging.

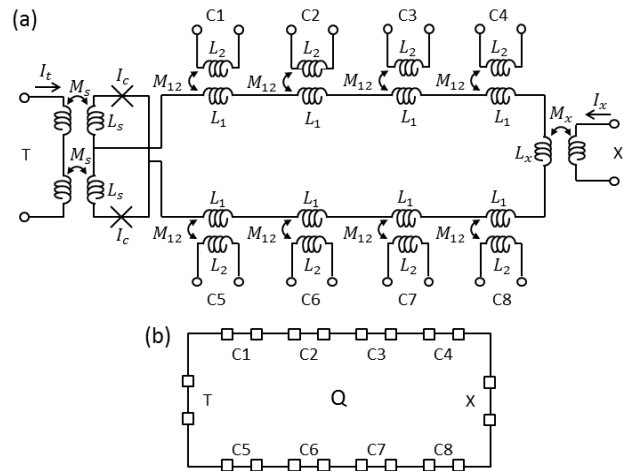

Fig. 3. A flux qubit for QA factoring. (a) schematic and (b) symbol. Ports, C1, C2, ... and C8, are used for interconnection with other qubits and auxiliary circuits. The qubit-bias and transvers fields are induced by currents $I_x$ and $I_t$, respectively.

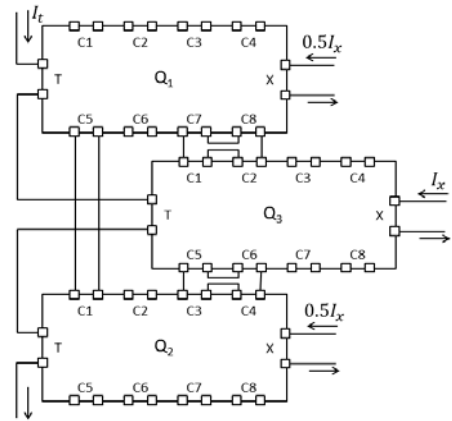

Fig. 4. Implementation of QA NOR with three superconducting qubits.

Fig. 5 shows a schematic cross-section of a three-dimensional packaging, a QUbit-chip / Interposer / Package-substrate (QUIP) structure, which we propose for realizing practical-scale QA factoring machines on ASAC architecture. A qubit chip, on which the qubits and couplers are densely integrated, is fabricated by using a modified version of our standard technology developed chiefly for superconducting digital circuits [15]. The modification includes the deposition of Nb/AlO$_x$/Nb Josephson-junction trilayers directly on hydrogen-terminated silicon substrates and the elimination of noisy disordered oxide such as anodized NbO$_x$, which are expected to improve the qubit performance. An active interposer, on which the qubit chips are mounted by using a flip-chip technology with superconducting solder bumps, includes the control and readout circuits consisting of moderate-size Josephson junctions. Superconducting through silicon vias are also fabricated in the active interposers for package interconnection. A package substrate, on which the active interposers with the qubit chips are packaged, is used for arranging a large number of wirings to connect the QA machine and room temperature electronics. In addition, bridge interposers are introduced for connecting the adjacent active interposers on the package substrate.

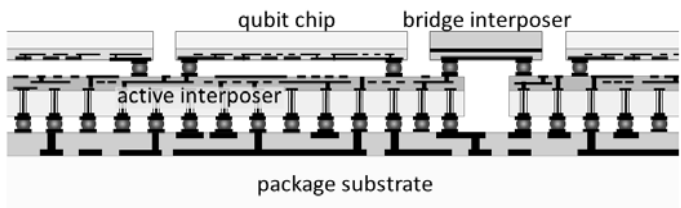

Fig. 5. QUbit-chip / Interposer / Package-substrate (QUIP) packaging for practical-scale QA machine.

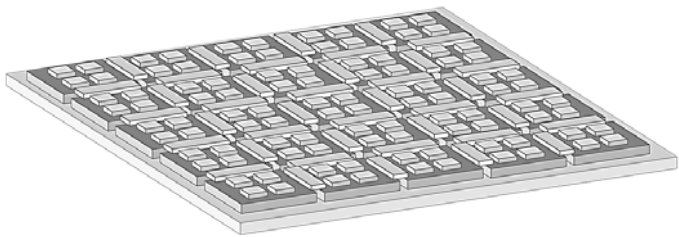

Fig. 6. A possible scheme toward integration of 6 million qubits for 512-bit factoring.

We estimate an occupation area per qubit to be 6400 $\mu m^2$ including the peripheral solder-bump bonding pads, suggesting a possible integration of 62 thousand qubits on a 20-mm-square chip. With these numbers, we simply expect factoring of a 512-bit number by a QA machine consisting of 100 qubit chips on a 250-mm-square package substrate (Fig. 6). Perhaps the number of bits to be factored, 512, is not very impressive but nevertheless the integration scale, 6 million qubits, is very challenging. Reduction of the number of the qubits is critical for constructing practical-scale QA machines for practical problems. Efficient designs of the QA Boolean logic circuits with the smaller number of qubits are discussed in detail elsewhere [17]. There also remain a lot of issues to be considered, *e.g.*, a method of calibrating the individual qubits, arrangements of the fine tuning mechanisms such as the CCJJ SQUID and Iqp compensator [8]-[10], and reduction of the number of cables between low and room temperatures.

## V. SUMMARY

We have designed a superconducting QA machine to perform prime factoring for analyzing RSA cryptosystem. The operation principle is the unitary reversibility of a multiplier composed of QA Boolean logic circuits. A new concept of an ASAC architecture efficiently utilizes the limited hardware budget in superconducting integrated circuit technology in exchange for the restriction of the functionality. A practical-scale ASAC QA factoring machine is to be realized in a three-dimensional QUIP packaging structure.

ACKNOWLEDGMENT

We are grateful to T. Endo, G. Fujii, M. Hioki, K. Inomata, T. Katashita, K. Kikuchi, S. Kohjiro, S. Nagasawa, T. Nakagawa, M. Ukibe, T. Yamada, and H. Yamamori for valuable discussions.